\documentstyle[11pt,ams,dina4]{article}
\begin{document}
\parindent0pt
\parskip1.5ex plus .6ex minus .3ex
%
%
%
%
\newcommand{\PICsnbif}{1}
\newcommand{\PICbifdiaP}{2a}
\newcommand{\PICbifdiaM}{2b}
\newcommand{\PICspketren}{3}
\newcommand{\PICKoeffSoft}{4}
\newcommand{\PICSkizzeSoft}{5}
\newcommand{\PICKoeffHardP}{6}
\newcommand{\PICKoeffHardM}{7}
\newcommand{\PICKoeffHardMO}{7a}
\newcommand{\PICKoeffHardMU}{7b}
\newcommand{\PICKoeffCodim}{8}
\newcommand{\PICKoeffCodimP}{8a}
\newcommand{\PICKoeffCodimM}{8b}
\newcommand{\PICSkizzeCodim}{9}
%
%
\newcommand{\dg}{$\delta$-$\gamma\ $}
\newcommand{\eqn}{eq.}
\newcommand{\Eqn}{Eq.}
\newcommand{\gs}{g}
\newcommand{\conjC}{{\cal C}}
\newcommand{\imG}{(i - \, \Gamma)}
\newcommand{\epGq}{1+\Gamma^2}
\newcommand{\epiG}{1+i\,\Gamma}
\newcommand{\emiG}{1-i\,\Gamma}
\newcommand{\omc}{\omega_c}
\newcommand{\voa}{v_{0a}}
\newcommand{\vob}{v_{0b}}
\newcommand{\woa}{w_{0a}}
\newcommand{\wob}{w_{0b}}
\newcommand{\po}{\Phi_0}
\newcommand{\poo}{\Phi_{00}}
\newcommand{\poq}{\bar\Phi_0}
\newcommand{\pooq}{\bar\Phi_{00}}
\newcommand{\pa}{\Phi_1}
\newcommand{\paq}{\bar\Phi_1}
\newcommand{\pb}{\Phi_2}
\newcommand{\pbq}{\bar\Phi_2}
\newcommand{\eppo}{1+\,|\Phi_0|^2}
\newcommand{\eppoo}{1+\,|\Phi_{00}|^2}
\newcommand{\pobq}{|\Phi_0|^2}
\newcommand{\poobq}{|\Phi_{00}|^2}
\newcommand{\be}{\begin{equation}}
\newcommand{\ee}{\end{equation}}
\newcommand{\bea}{\begin{eqnarray}}
\newcommand{\eea}{\end{eqnarray}}
\renewcommand{\Re}{\mbox{Re}\,}
\renewcommand{\Im}{\mbox{Im}\,}
\renewcommand{\O}{\script{O}}
%
%
\title{Amplitude equations near pattern forming instabilities
       for strongly driven ferromagnets}

\author{F.~Matth\"aus and H.~Sauermann\\
        Technische Hochschule Darmstadt\\
        Institut f\"ur Festk\"orperphysik\\
        Hochschulstra\ss{}e 8\\
        D--64289 Darmstadt\\
        Germany\\
        Tel.: + 49 6151 165395\\
        FAX   + 49 6151 164165\\
        e-mail: frank@arnold.fkp.physik.th-darmstadt.de}

\date{January 19, 1995}
\maketitle

%
%
\begin{abstract}
A transversally driven isotropic ferromagnet being under the
influence of a static external and an uniaxial internal anisotropy
field is studied.
We consider the dissipative Landau-Lifshitz equation
as the fundamental equation of motion and treat it in $1+1$~dimensions.
The stability of the spatially homogeneous magnetizations
against inhomogeneous perturbations is analyzed.
Subsequently the dynamics above threshold is described via
amplitude equations and the dependence of their
coefficients on the physical parameters of the system is
determined explicitly.
We find soft- and hard-mode instabilities, transitions
between sub- and supercritical behaviour,
various bifurcations of higher codimension, and
present a series of explicit bifurcation diagrams.
The analysis of the codimension-2 point where the
soft- and hard-mode instabilities coincide leads to
a system of two coupled Ginzburg-Landau equations.
\end{abstract}
\newpage
%
%
\section{Introduction}

By now there is a huge amount of work treating pattern formation
in magnetically ordered substances under equilibrium conditions,
i.e. under the influence of static external as well as internal fields.
It has led to
what seems to be
both theoretically and experimentally
a rather complete picture.
The situation is much more precarious if the system is driven
far from equilibrium by applying strong oscillating magnetic fields.
Being hailed as a canonical example for pattern formation
by some authors (comp. e.g. \cite{Ande1981}), critical
questions have been raised by others \cite{CrHo1993}.

The treatment of ferromagnetic resonance phenomena,
which has been initiated by Suhl~\cite{Suhl1957} and uses
spin waves and their nonlinear interactions, has proved to be an important
development in this field.
Truncating the number of relevant modes
properly many results have been obtained which
explain most of the observed phenomena pretty successfully.
They comprise the occurrence of threshold values
for parametric instabilities as well as
typical nonlinear effects including
various bifurcation scenarios to low dimensional chaos.

Even the use of the full (nontruncated) set of spin wave equations
is doubtful, however, if one discusses pattern formation for
large aspect ratios in spatially extended systems, because
it lacks a systematic derivation.
Consequently Elmer~\cite{Elme1987} has proposed another approach.
Starting from the fundamental Landau-Lifshitz equation of motion
describing the dynamics of the magnetization field on a mesoscopic level
and including various damping mechanisms (see below)
he derived,
for fixed values of the system parameters,
amplitude equations by using multiple-scale perturbation theory.
His main objective was to discuss his findings as functions
of the strength of the driving field.

We choose as our starting point the dissipative Landau-Lifshitz
equation
\be\label{eqn:LL}
 \dot{\vec{S}}(x,t)  =   - \left( \vec{S} \times \vec{H}_{eff} +
      \Gamma \ \vec{S} \times \left( \vec{S} \times
      \vec{H}_{eff} \right) \ \right).
\ee
containing a Gilbert damping term on the right hand side as well.
The latter has been justified microscopically in~\cite{Gara1991,Plef1993}.
Disregarding the dipolar interaction as far as it cannot be incorporated
into a local anisotropy field we study it adopting a rather
different point of view.
Based on a work by Tr\"axler et al.~\cite{TTJS1995},
where it has been shown that the homogeneous
solutions of equation~(\ref{eqn:LL})
possess a surprisingly rich bifurcation scenario when
the physical parameters of the system
are varied, we extend their work by admitting spatial inhomogenities.
As in their article we keep the amplitude of the driving field fixed
and use the static external field together with the detuning
from resonance as fundamental parameters.
Aiming at obtaining a complete overview in the space of parameters
we find sub- and supercritical soft-mode and, for the first
time for this magnetic system, hard-mode instabilities also.
The corresponding amplitude equations are derived together with
all of their coefficients explicitly.

We then proceed to show that our results imply the existence of some
what we deem to be highly interesting bifurcations of higher codimension.
We just mention here the codimension-2 example discussed in section~5.3
which arises when the soft- and hard-mode instabilities coincide
and which leads to a system of two coupled real and complex
Ginzburg-Landau equations.

However, in our opinion the results given here represent a contribution
to earning strongly driven ferromagnets a solid place among
the classic pattern forming forebears in
hydrodynamics, nonlinear optics, etc.~\cite{CrHo1993}.

%
%
\section{Equation of motion}

We start from the dissipative Landau-Lifshitz equation
of motion~(\ref{eqn:LL}) and assume that
the effective field $\vec{H}_{eff}$ is composed by
external fields, i.e.
a static field $h_z$ and a circularly polarized driving field $h_\perp$
as well as internal fields.
The latter are an uniaxial anisotropy field $a\,S_z$ and an isotropic
exchange field ($J>0$).
\be
 \vec{H}_{eff}  = \ ( h_z + a \ S_z ) \  \vec{e}_z +
      J \ \left( \frac{\partial^2 \vec{S}}{{\partial x}^2}\  \right) +
      h_\bot \ ( \cos{\omega t} \ \vec{e}_x + \sin{\omega t} \ \vec{e}_y ).
\ee
To get rid of the explicit time dependence we first transform to
a rotating frame of reference,
a fact which must not be forgotten when interpreting our results.
\be\label{eqn:RotFrame}
 {S}_x^\prime = S_x \cos(\omega\,t) + S_y \sin(\omega\,t),\
S_y^\prime = -S_x \sin(\omega\,t) + S_y \cos(\omega\,t), \ S_z^\prime = S_z.
\ee
Observing then that \eqn~(\ref{eqn:LL})
preserves the modulus of the local spin density $\vec{S}(x,t)$,
we may put $|\vec{S}|=1$.
Eliminating the superfluous coordinate with the help
of a stereographic projection
\be
  \Phi = \frac{S_x^\prime + i\, S_y^\prime}{1+S_z^\prime}
\ee
we obtain the final form for the equation of motion
\bea\label{eqn:LLstereo}
   \frac{\partial \Phi}{\partial t} & = & \  \
    (i-\Gamma) \ \left[ (\delta + i\,\gamma) \, \Phi +
              a \,\Phi\ \left( \frac{1-|\Phi|^2}{1+|\Phi|^2} \right) -
              \frac{h_\bot}{2} \left( 1 - \Phi^2 \right) \right.\\ \nonumber
   & &  \hspace{1.7cm}
        \left. + \,J \,\left( \frac{2\,\overline{\Phi}} {1+|\Phi|^2} \
         {\left( \frac{\partial \Phi}{\partial x} \right)}^2
         - \frac{\partial^2 \Phi}{{\partial x}^2 } \right) \right].
\eea
The parameters
\be \delta = h_z - \frac{\omega}{1+ \Gamma^2}
      \hspace{1cm} \mbox{and} \hspace{1cm}
   \gamma = \frac{\omega \, \Gamma}{1+ \Gamma^2}
\ee
denote the detuning from resonance and the renormalized
external frequency, respectively.

Considering the set of parameters characterizing
the effective field it seems
reasonable to scale them in terms of the anisotropy $|a|$.
Thus $a$ takes on only the values $\pm 1$,
distinguishing between the easy-axis ($+1$)
and easy-plane ($-1$) case.
Furthermore in \eqn~(\ref{eqn:LLstereo})
the exchange constant $J$ may be absorbed into the
variable $x$.

Studying the bifurcation behaviour of this system we may confine ourselves
to the case $h_\perp>0$ and $\gamma>0$.
A negative value of $h_\perp$ merely corresponds to
shifting the phase of the driving field by $\pi$.
Furthermore \eqn~(\ref{eqn:LLstereo})
remains invariant under the transformation
($\Phi \rightarrow \frac{1}{\Phi}, \delta \rightarrow -\delta,
\gamma \rightarrow -\gamma$), implying
that all bifurcation diagrams are symmetric with respect to the
origin in the \dg plane.

Our analytical calculations of bifurcation lines, amplitude equations, etc.
are performed without imposing any restrictions on the parameters
$\,\Gamma$, $h_\perp$, $\delta$ and $\gamma$.
For numerical purposes we will always choose $\Gamma=0.1$ and keep
$h_\perp$ fixed at $h_\perp=0.1$.
This means that we investigate the bifurcation behaviour in the remaining
\dg parameter plane.
For the \mbox{codimension-2} bifurcation, to be discussed later on,
the variation of the coefficients
of the amplitude equations will be given as a function of $h_\perp$ also.

%
%
\section{Spatially homogeneous fixed point solutions}

We start our systematic investigation of \eqn~(\ref{eqn:LLstereo})
by analyzing its spatially homogeneous fixed point solutions
$(\, \partial_x \Phi_0 = \partial_t \Phi_0 \, = 0)$.
A rather complete bifurcation analysis
-- including codimension-3 bifurcations --
of the spatially homogeneous system
can be found in~\cite{TTJS1995}.
Thus we confine ourselves to a description of some elementary results
which are important in the present context.
The spatially homogeneous
time independent solutions of \eqn~(\ref{eqn:LLstereo})
for example are shown to be given by the roots of the algebraic equation:
\be\label{eqn:fpSz}
 (\gamma^2+a^2)\,S_z^4 + 2\,a\,\delta\ S_z^3 +
 (\delta^2+h_\perp^2-a^2-\gamma^2)\ S_z^2 -
 2\,a\,\delta\ S_z - \delta^2 = 0.
\ee
Here $S_z$ is the cartesian spin coordinate.
{}From the solution of \eqn~(\ref{eqn:fpSz}) $S_x$ and $S_y$
(and therefore $\Phi$) may be determined explicitly.
Depending on the values of the parameters
this polynomial has either two or four solutions.
The corresponding regions in parameter space
are separated by saddle--node bifurcation lines as
is shown in figure~\PICsnbif.
\marginpar{$< Fig. 1$}
As the substitution $a \rightarrow -a$ and  $S_z \rightarrow -S_z$
leaves \eqn~(\ref{eqn:fpSz}) invariant, this bifurcation does not
depend on the type of anisotropy chosen.
Although the saddle-node bifurcations can be calculated easily from
\eqn~(\ref{eqn:fpSz}) we skip this calculation here because an
equivalent expression is obtained within the framework
of our stability analysis in the next section.

%
%
\section{Stability analysis}

Let $\Phi_0$ denote a spatially homogeneous fixed point solution
of \eqn~(\ref{eqn:LLstereo}).
Proceeding as usual in discussing linear stability
we add an infinitesimal perturbation $\delta\Phi(x, t)$
\be\label{eqn:LinStabDef}
 \Phi(x, t) = \Phi_0 + \delta\Phi(x, t).
\ee
Its time evolution is governed in first order
by the equation
\be\label{eqn:LinStab}
 \delta\dot{\Phi} = (i - \, \Gamma) \,
      \left[ \gs_1 - \partial_x^2 + i \,\gs_2 +
            (\gs_3 + i \,\gs_4) \, \conjC \right] \, \delta\Phi,
\ee
where $\conjC$ denotes the operator of complex conjugation,
i.e. $\conjC f = \bar{f}$
and $\gs_1, \ldots, \gs_4$ are given by
\bea\label{eqn:gammas}
  \gs_1 & = & \delta + h_\perp \Re(\Phi_0) +
      a \ \frac{1- 2 \ | \Phi_0 |^2 - | \Phi_0 |^4}
             {(1+| \Phi_0 |^2)^2} \\
  \gs_2 & = &  \gamma + h_\perp \, \Im(\Phi_0) \\
  \gs_3 & = & \frac{-2 \ a}{(1+| \Phi_0 |^2)^2}
                 \ \Re(\Phi_0^2) \\
  \gs_4 & = & \frac{-2 \ a}{(1+| \Phi_0 |^2)^2}
                 \ \Im(\Phi_0^2).
\eea
The plane wave solutions of \eqn~(\ref{eqn:LinStab})
are written in the form
\be\label{eqn:LinMode}
 \delta\Phi_q = \cos(q x) \,
   \left( \delta\Phi_+ \ e^{\lambda t} +
          \delta\Phi_- \ e^{\bar\lambda t} \right),
\ee
which accommodates for
the presence of the complex conjugation operator.
This yields after some algebra the following secular equation
\be\label{eqn:secular}
  \lambda^2 - \lambda \ tr(q) + det(q) = 0,
\ee
with
\bea
  tr\,(q)  & = &
     -2 \ \left( \gs_2 + \Gamma \ (\gs_1 + q^2) \right)
  \label{eqn:tr} \\
  det\,(q) & = & (1+\, \Gamma^2)\,
  \left[ (\gs_1 + q^2)^2 + \gs_2^2 - \gs_3^2 - \gs_4^2 \right].
  \label{eqn:det}
\eea
Here $tr$ and $det$ denote the trace and determinant of
the two by two $q$--dependent coefficient matrix
which arises when \eqn~(\ref{eqn:LinStab}) is split
into real and imaginary part.
As is obvious from their
definitions $\{\gs_\alpha\}$,
$tr$ and $det$ depend
on the fixed point coordinates $\Phi_0(\delta, \gamma,h_\perp,a=\pm 1)$
as well as on the parameters~($\delta,\gamma,h_\perp,a=\pm 1,\Gamma$)
and on the wavenumber~$q$.

The solution $\Phi_0$ is linearly stable if all possible
perturbations~$\delta\Phi_q$ decay exponentially in time, i.e.
if $\Re \lambda_\pm(q) < 0$ for all $q$,
where $\lambda_\pm(q)$ denotes the two roots of \eqn~(\ref{eqn:secular})
for some fixed wavenumber~q.
As is described in~\cite{KuTs1975,ToOT1974}
this yields the conditions
\be\label{eqn:TrDetCond}
  tr\,(q) < 0 \hspace{1.2cm}  \mbox{and} \hspace{1.2cm}
  det\,(q) > 0
\ee
which have to be fulfilled simultaneously for all $q$.
Thus there are two different possibilities for these
stability conditions to become violated,
leading to soft-mode and hard-mode instabilities.
When both of them are violated simultaneously we have a codimension-2
bifurcation which requires a separate discussion
to be given later on.

\subsection{Soft-mode instability}

A soft-mode instability is characterized by
a zero eigenvalue $\lambda_+(q_c)=0$ which happens if
$det(q_c)$ vanishes. More precisely it occurs if the minimum
value of $det(q)$ becomes zero for some finite $q=q_c$,
while $tr(q)$ remains negative.
Hence we get from \eqn~(\ref{eqn:det})
\be\label{eqn:soft:dqdet}
   \partial_q det|_{q=q_c} = (\epGq)\ 4\,q_c\ (\gs_1 + q_c^2) = 0
\ee
\be
   det|_{q=q_c} = (\epGq)\  \left( (\gs_1 + q_c^2)^2 +
        \gs_2^2 - \gs_3^2 - \gs_4^2\right) = 0.
\ee
Thus for $q_c \neq 0$ we end up with the bifurcation condition
\be\label{eqn:BifBed:Soft}
  \gs_2^2 - \gs_3^2 - \gs_4^2 = 0
\ee
defining a surface in the space of parameters.
Its boundary results from
$-\gs_1 = q_c^2 = 0$, which leads to
the end points of the bifurcation
lines in the \dg plane for $\Gamma$ and $h_\perp$ fixed.

The second solution $q_c=0$ of~\eqn~(\ref{eqn:soft:dqdet})
just represents the spatially homogeneous saddle--node
bifurcation mentioned in section~3.
It is described by
$\gs_1^2 + \gs_2^2 - \gs_3^2 - \gs_4^2 = 0$.
The corresponding bifurcation lines have been depicted already
in figure~\PICsnbif.

\subsection{Hard-mode instability}

In the case of a hard-mode instability
the eigenvalues $\lambda(q_c)_\pm=\pm i\,\omega(q_c)$ become
purely imaginary at the instability point.
This occurs if $tr(q_c)$, more precisely its maximum value
at some $q=q_c$, vanishes, while
$\omega_c^2= det(q_c)$ remains positive.
{}From \eqn~(\ref{eqn:tr}) we get immediately
\be
   \partial_q tr|_{q=q_c} = -4 \ \Gamma\ q_c = 0, \hspace{10mm}
       i.e. \hspace{4mm}q_c = 0
\ee
\be\label{eqn:BifBed:Hard}
   tr|_{q_c=0} = -2\,\left(\gs_2 + \Gamma \ \gs_1\right) = 0.
\ee
Hence the bifurcation manifold is given by
\eqn~(\ref{eqn:BifBed:Hard})
and its boundary by $\omega_c^2= det = 0$.

The form of \eqn~(\ref{eqn:tr}) implies
that the critical modes are spatially homogeneous.
Thus there are no primary instabilities creating
travelling wave states ($\omega_c\neq 0, q_c\neq 0$).

\subsection{The bifurcation diagram}

To obtain the soft- and hard-mode as well as saddle-node
bifurcation lines in the space of the physical parameters
we proceed as follows:
Keeping $\Gamma=0.1, a=\pm1$, $h_\perp=0.1$ fixed,
we solve the
two (real) fixed point equations --for convenience we use
\eqn~(\ref{eqn:LLstereo}),
and not the polynomial~(\ref{eqn:fpSz})--
together with the proper bifurcation condition numerically.
Using the {\sc Pitcon}-package~\cite{Pitcon} we track
the solution curves of this algebraic problem
in the four dimensional
space spanned by $\Re(\Phi_0), \Im(\Phi_0), \delta$ and $\gamma$.
Its projections onto the \dg plane are the
bifurcation lines sought for.
For the special case of the spatially homogeneous
saddle-node bifurcation the result has been presented
already in figure~\PICsnbif.
The soft- and hard-mode instabilities
\marginpar{\mbox{$< Fig. 2$}}
which complete this diagram
are drawn in figures~\PICbifdiaP{}  and \PICbifdiaM{}
for both types of anisotropy.
The dotted parts of the bifurcation lines
indicate that the corresponding instability occurs only
after the solution $\Phi_0$ being concerned
has been destabilized already by another type of instability.
It is important especially for the hard-mode
instability to keep this physically ``hidden'' part
of the bifurcation line in mind when discussing
some of the results which are related to the amplitude
equations.

The gross features\footnote{See reference~\cite{TTJS1995}
for more details concerning the
spatially homogeneous system.}
of the solutions of our system following from
figures \PICbifdiaP{} and \PICbifdiaM{} may be summarized as follows:

In the case of an easy-axis anisotropy (cf.~fig.~\PICbifdiaP{})
we find in region (I) of the parameter plane a stable solution
near the north pole ($\Phi=0$)
and an unstable solution near the south pole ($\Phi=\infty$).
Crossing the hard-mode bifurcation line from region (I)
to region (II)
the southern fixed point is stabilized and an unstable
spatially homogeneous limit-cycle emerges.
This limit-cycle in turn destabilizes the northern fixed point solution
at the second hard-mode bifurcation line.
Thus in region (III) we have a stable solution
on the southern and an unstable solution on the northern hemisphere.
Crossing the soft-mode bifurcation line
separating regions (II) and (IV)
the northern (stable) spatially homogeneous
fixed point solution gets destabilized with regard to
spatially periodic states.

For an easy-plane anisotropy (cf.~fig.~\PICbifdiaM{}) the situation
in regions (I) and (III) is the same as in
the easy-axis case. In region (II), however,
both homogeneous fixed point solutions are unstable.
The northern fixed point of region (I) is now
affected by the upper hard-mode line
but not by the saddle-node bifurcation ($\overline{76}$).
It looses stability with regard to spatially periodic perturbations
when passing the soft-mode line separating regions (Ia) and (VI).
Note that in accordance with the stability properties of the fixed point
involved in the pitchfork bifurcation~(6) two stable homogeneous states exist
in region (Ia).

In this survey we have neither mentioned the fact, that on the
lower {\sc Hopf} line a transition from super- to subcritical
bifurcations occurs implying saddle--node bifurcations
of limit--cycles
nor referred to the global bifurcations of the homogeneous
system found in~\cite{TTJS1995}.

We stress, however, that our analysis leads via \eqn~(\ref{eqn:soft:dqdet})
to an explicit expression for $q_c$ along the soft-mode line.
Starting from its origin at the homogeneous saddle-node bifurcation
line $q_c$ turns out to increase monotonously from
zero to infinity.
The corresponding results for $\omega_c$ are contained in~\cite{TTJS1995}
already.

We finally present in figure~\PICspketren{} a number of
typical spectra for several fixed points at salient
positions in the \dg plane.
\marginpar{$< Fig. 3$}

%
%
\section{Amplitude equations\label{sec:agln}}

After having calculated the possible instabilities
and the corresponding bifurcation lines
in the space of parameters, we go on by analyzing
the behaviour of our system
in the neighbourhood of these marginal stability lines.
This will be achieved with the help of a perturbation expansion.
A natural expansion parameter
$\epsilon$ measuring the distance from threshold
results from writing
\bea
  \delta & = & \delta_c + \epsilon^2\,\delta_2 \\
  \gamma & = & \gamma_c + \epsilon^2\,\gamma_2.
\eea
($\delta_c, \gamma_c$) are the parameters at criticality
whereas ($\delta_2, \gamma_2$),
which may be thought of as the components of a unit vector,
characterize the direction in which the bifurcation line
is crossed transversally.

Let quite generally $u_0(x,t)$ denote the critical mode at instability.
In order to apply perturbation theory above threshold
we put as usual
\be\label{eqn:loesungsansatz}
  \Phi = \Phi_0 + \epsilon\,A(\epsilon\,x,\epsilon^2\,t)\,u_0(x,t)+
      \epsilon^2\,r(x,t,\epsilon).
\ee
The scaling of the slow space and time coordinates
\be
 X = \epsilon \ x \hspace{1.2cm} \mbox{and} \hspace{1.2cm} T = \epsilon^2 t
\ee
is motivated
by the quadratic dependence of the spectrum on the wavenumber~$q$
near criticality.
The slowly varying amplitude $A(X, T)$ is determined so
that the solution~(\ref{eqn:loesungsansatz}) contains no secular terms.
Inserting the ansatz~(\ref{eqn:loesungsansatz}) into the equation
of motion~(\ref{eqn:LLstereo}) this requirement leads via
solvability conditions in lowest non-trivial order of perturbation theory
to an equation governing the time evolution of $A$.
This procedure is well known and has been often applied especially
to hydrodynamic problems (cf. ref.~\cite{NeWh1969,Sege1969}).
For a review see the extensive
paper by Cross and Hohenberg~\cite{CrHo1993}
as well as the book by Manneville~\cite{Mann1990}.
Kuramoto and Tsuzuki~\cite{KuTs1975,Kura1984} used this technique
in connection with reaction--diffusion equations and derived
the coefficients of the amplitude equations for a large
class of fundamental equations exhibiting
soft- and hard-mode instabilities.
Our equation of motion~(\ref{eqn:LLstereo}) does not fit into
their scheme because of the quadratic derivative
$(\frac{\partial\Phi}{\partial x})^2$. It will affect the nonlinear
coefficient of the soft-mode amplitude equation.

Concerning magnetic systems this procedure was first applied
by Elmer in the context of ferromagnetic resonance.
His early work is based on a Landau-Lifshitz equation containing a
dipolar interaction term whereas dissipative effects
are taken into account employing the
Bloch-Bloembergen damping mechanism~\cite{Elme1987}.
More recently he described pattern formation
in thin magnetic films studying
a Landau-Lifshitz equation with Gilbert damping~\cite{Elme1993}.
In contrast to our work he is interested mainly
in the effects caused by the dipolar interaction.

The structure of the amplitude equations is determined
completely by the nature of the instability considered
and the symmetries of the underlying system.
They are thus universal, usually turn out to be
real or complex Ginzburg--Landau
equations and are found in a wide variety of physical systems
(see for example the numerous references in~\cite{CrHo1993}).
The individual physics of the system at hand
manifests itself in the explicit values of their coefficients,
more precisely in their dependence on the physical parameters.
We aim at giving a fairly complete
survey in this matter for our magnetic problem.

In the following section we present our results for the
amplitude equations
for the various instabilities and plot the variation of
their coefficients along the corresponding bifurcation lines.
As the necessary calculations in deriving these coefficients
explicitly are very lengthy they are deferred to appendix~A.

It has been pointed out (see ref.~\cite{CrHo1993} and the
references therein) that the linear coefficients
of the amplitude equations
(cf.~\eqn(\ref{eqn:agl:softmode},\ref{eqn:agl:hardmode},\ref{eqn:agl:codim}))
can be derived
alternatively by differentiating the spectrum of the
linearized equation of motion.
In our case this amounts to
--cf.~\eqn~(\ref{eqn:LinStab},\ref{eqn:secular})--
\bea
  \label{eqn:LinMuGen}
  \mu &=& \left.\frac{\partial \lambda}{\partial \gamma}\right|_c
          \, \gamma_2 +
          \left.\frac{\partial \lambda}{\partial \delta}\right|_c
          \, \delta_2 \\
  \label{eqn:LinAlphaGen}
  \alpha & = & -\,\frac{1}{2}
             \left.\frac{\partial^2 \lambda}{{\partial q}^2}\right|_c.
\eea
The subscript means that these expressions have to be calculated
at the critical values.
Note that the linear coefficient $\mu$ depends on the
direction in which the bifurcation line in the \dg
plane is crossed when entering the unstable region.
We will always choose it to be normal to that line.
The diffusion coefficient $\alpha$ marks the dissipative
part of the amplitude equation, whereas the nonlinear
coefficient $r$ characterizes the type of bifurcation,
i.e. for example whether it is super- or subcritical, and
fixes the saturation value of the amplitude in the supercritical case.

%
%
\subsection{Soft-mode instability}

For a soft-mode instability the slowly
varying amplitude $A(X, T)$ has been introduced in appendix~A
according to
\be
  \Phi - \Phi_0 = \epsilon\,v_0 \left(
      A(\epsilon\,x,\epsilon^2\,t)\,e^{i\,q_c\,x} + c.c. \right).
\ee
Its time evolution is found to be governed by
a Ginzburg-Landau equation~\cite{KuTs1975,NeWh1969,Sege1969}
with real coefficients:
\be\label{eqn:agl:softmode}
  \partial_T\,A = \mu_a\,A + \alpha_a\,\partial_X^2\,A - r_a \ |A|^2\,A.
\ee
This equation possesses a Lyapunov potential and
is therefore of relaxational type.
As the modulus of all
of its coefficients may be rescaled to unity by appropriate
redefinitions of time, space and amplitude,
the most interesting traits of this equation
show up in situations where these coefficients change sign
and as a consequence the behaviour of the solutions
changes qualitatively.

The variation of the coefficients along the
bifurcation line is shown in figure~\PICKoeffSoft{}
for the easy axis case.
\marginpar{\mbox{$<Fig.~\PICKoeffSoft$}}
The diffusion coefficient $\alpha_a$, which is proportional
to $q_c^2$, is zero at its origin which lies
on the saddle-node bifurcation line and increases
monotonously.
At the origin itself
the derivation of the amplitude equation looses its validity,
an issue which will be discussed separately.
Notice that the third-order coefficient changes sign
along the bifurcation line signalling a transition from a
subcritical to a supercritical bifurcation.
In the supercritical case we have a (forward) bifurcation of
a stable spatially periodic state, whereas for subcritical
behaviour the system shows a (backward) bifurcation
of an unstable spatially periodic state.
Whether or not starting from a perturbation theoretical treatment
in a neighbourhood of this degenerated point higher order terms
lead to a stabilization remains undecided.
In view of appendix~A such calculations seem to be
impracticable however, so that probably one will have to
resort the full \eqn~(\ref{eqn:LLstereo})
again to answer such questions. Work in this
direction will be reported.

In order to achieve some qualitative understanding of
this transition we have
performed direct numerical simulations on the
equation of motion~(\ref{eqn:LL}) using parameter values
in the neighbourhood of this point.
\marginpar{$<Fig.~\PICSkizzeSoft$}
Crossing the soft-mode bifurcation line
at point~A (cf. fig.~\PICSkizzeSoft{}) in the supercritical region ($r>0$)
a periodic state emerges continuously.
Pursuing that solution along the curve {\cal C}
we find that it continues to exist with finite amplitude
when point B on the subcritical side is reached.
So, beyond point B a stable homogeneous and
a stable periodic solution coexist.
We conjecture that --in close analogy to degenerated {\sc Hopf}
bifurcations-- there is a line of saddle node bifurcations
of spatially periodic solutions on which the stable periodic
solution is destroyed together with the unstable periodic
one which is created when crossing the
soft-mode bifurcation line at point~B.

We have not plotted the coefficients as functions of the parameters
for negative anisotropies along the segment of the soft-mode bifurcation
line connecting the (homogeneous) saddle-node and the
hard-mode bifurcation lines
because this bifurcation is subcritical throughout.
Recall that all these lines affect one and the same solution.
Numerical simulations reveal that
the solution merely relaxes towards the other stable
(spatially homogenous) state after a transient period during
which periodic structures appear.

The obvious divergences of the linear and nonlinear coefficients
$\mu$ and $r$ are rooted in the fact that the critical
wavenumber~$q_c$ vanishes, when the
soft-mode bifurcation approaches the saddle-node bifurcation.
Firstly, the fixed point equations,
depending on the parameters $\delta$ and $\gamma$,
have to be expanded with respect to $\epsilon^2$
in a neighbourhood of the critical parameter values.
This can be done consistently only if the corresponding
Jacobian has non-zero eigenvalues.
At a saddle-node bifurcation this
condition is violated and the expansion breaks down
leading to a diverging behaviour for the coefficient $\mu$.
Secondly, as far as the nonlinear coefficient $r$ is concerned,
we have to solve an equation second order in $\epsilon$.
(Refer to appendix~A for details).
The corresponding solution looses its validity however for
vanishing critical wavenumber because
the linear operator on the left hand side simply
reduces to the Jacobian so that trivially
the solvability condition is no longer satisfied.
Thus at this point in parameter space the formalism
used to derive the amplitude equation becomes invalid.
A similar situation has been observed by
Kuramoto et. al. in~\cite{KuTs1975}.

%
%
\subsection{Hard-mode instability}

The amplitude equation for this type of instability
is given by
\be\label{eqn:agl:hardmode}
  \partial_T\,B = (\mu_{bR} + i\,\mu_{bI})\,B +
                  \alpha_{b}\,(1+ i\,c_1)\,\partial_X^2\,B
                 - r_{b}\,(1 - i\,c_2) \ |B|^2\,B.
\ee
In contrast to the soft-mode case this
equation has complex-valued coefficients.
The real part $\alpha_{b}$ of the diffusion coefficient
is identical to the damping constant $\Gamma$
as is proved in appendix~A, \eqn~(\ref{eqn:diffconst:hard}).
The imaginary part $\mu_{bI}$ of the linear coefficient
can be removed by a transformation
to a rotating frame ($B \rightarrow B\,e^{i\,\mu_{bI}\,t}$).

This complex Ginzburg-Landau equation has been addressed to frequently
in the literature and is analysed in a wide variety of contexts.
The behaviour of its solutions ranges from relaxational dynamics
for vanishing imaginary parts of the coefficients ($c_1,c_2 =0$)
over some very complex -- even chaotic -- types
of spatio-temporal patterns
in an intermediate range of ($c_1, c_2$) to the completely
integrable limit ($c_1,c_2 \rightarrow \infty$)
of the nonlinear Schr\"odinger equation.

For easy-axis anisotropy the coefficients are plotted
as functions of $\delta$ along the whole lower hard-mode
bifurcation line (cf. figure~\PICbifdiaP{})
in figure~\PICKoeffHardP{}.
\marginpar{$< Fig.~\PICKoeffHardP$}
The singularities appearing when the bifurcation line
approaches the saddle-node bifurcation are caused by the vanishing
of the critical frequency~$\omega_c$.
Their explanation is completely analogous to that in the
soft-mode case for vanishing $q_c$ in the last section.
Regarding the coefficient $r_b$
we recognize a change of sign signalling a transition from
a subcritical to a supercritical bifurcation.
Unfortunately this degeneracy is found in a region of parameter space,
where the corresponding solution $\Phi_0$ is already unstable
due to the soft-mode instability
separating regions (II) and (IV).
If the dependence on the driving field $h_\perp$
is taken into account additionally, this situation changes as
will be expounded
in the following section in which the codimension-2
bifurcation is treated.

As has been demonstrated in~\cite{TTJS1995} for the spatially
homogeneous system explicitly, and is well known under
general circumstances~\cite{GuHo1983},
such degeneracy of a hard-mode bifurcation entails
a saddle-node bifurcation of (spatially homogeneous) limit-cycles.
The resulting bifurcation line ends at that point of the
hard-mode instability where the coefficient $r_b$ vanishes.

Along the upper hard-mode line (cf. fig.~\PICbifdiaP{})
the bifurcation is subcritical everywhere.
Hence the variation of the coefficients is not shown explicitly.
The analysis of the spatially homogeneous system~\cite{TTJS1995}
makes evident that this subcritical bifurcation leads to the unstable,
spatially homogeneous limit-cycle existing in region~(II) of
parameter space.

Concerning the easy-plane anisotropy, the variation of the coefficients
for both hard-mode lines is depicted in figure~\PICKoeffHardM{}.
\marginpar{\mbox{$<Fig.~\PICKoeffHardM{}$}}
Along the upper curve (cf.~fig.~\PICKoeffHardM{}a)
the bifurcation is supercritical
for all values of the parameters whereas along the
lower bifurcation line we recognize a transition from a sub-
to a supercritical bifurcation (cf.~fig.~\PICKoeffHardM{}b).
Regarding the supercritical portion of that line, the small numerical
value of the coefficient $r_b$ implies a large saturation amplitude
of the oscillatory solution.
We surmise that this is brought about by the stereographic
projection technique,
which leads to a large modulus of
$\Phi_0$ if the fixed point in question
lies on the southern hemisphere.
It is thus an artificial product.

%
%
\subsection{Codimension--2 bifurcation}

We now focus on the codimension-2 point in parameter space
where the hard- and soft-mode bifurcations fall together.
The spectrum of the linear operator at this point is depicted
in figure~\PICspketren{} (1) and indicates that the
degenerated critical mode
is now composed of a spatially periodic time-independent
and a time-periodic spatially homogeneous
part. Supplying both of them with
slowly varying amplitudes $A(X,T)$ and $B(X,T)$,
respectively, we prove in appendix~A.3 that
their dynamics is governed by two
coupled (real and complex) Ginzburg--Landau equations
\bea\label{eqn:agl:codim}
  \partial_T\,A &=& \mu_a\,A + \alpha_a\,\partial_X^2\,A
                     - r_a \ |A|^2\,A - s_a \,|B|^2\, A  \\ \nonumber
  \partial_T\,B &=& (\mu_{bR} + i\,\mu_{bI})\,B +
                    \alpha_{b}\,(1+ i\,c_1)\,\partial_X^2\,B
                    - r_{b}\,(1 - i\,c_2) \ |B|^2\,B \\ \nonumber
   & &              - (s_{bR} + i\,s_{bI}) \ |A|^2\,B.
\eea
The linear coefficients
as well as the nonlinear coefficients $r_a$, $r_b$ and $c_2$
are the same as those in the uncoupled equations
for the corresponding instabilities.
This is obvious from continuation arguments.
Only the coupling coefficients $s_a$ and $s_b=s_{bR}+i\,s_{bI}$
turn out to be qualitatively new quantities.
To get an impression of their behaviour
we study their functional dependence of
the amplitude $h_\perp$ of the transversal driving field
which was fixed up to now
(cf. fig.~\PICKoeffCodim{}).
\marginpar{\mbox{$<Fig.~\PICKoeffCodim{}$}}
We desist from discussing the linear coefficients $\mu_a$ and $\mu_b$
because it is inevitable to pay attention to the actual
direction in which the unstable region is entered
in this case (cf. fig.~\PICSkizzeCodim{}).
This was unnecessary hitherto, because the direction normal to the
bifurcation line has been a natural choice.

For both values $\pm1$ of the anisotropy we recognize
that the sign of the coefficient $r_{b}$ changes for
a certain amplitude $h_\perp$.
Thus for this value, giving rise to a bifurcation of even higher
codimension, the point of degeneracy of the hard-mode bifurcation,
separating its sub- and supercritical forms,
crosses the codimension-2 point where the hard- and soft-mode
instabilities coincide.

For still higher amplitudes $h_\perp$ -- and easy-axis anisotropy ($a=+1$)--
an interesting and complex situation arises
which is illustrated schematically
in figure~\PICSkizzeCodim{}.
\marginpar{\mbox{$<Fig.~\PICSkizzeCodim{}$}}
The degenerate hard-mode bifurcation
D enters region (II-III, fig.~\PICbifdiaP{})
so that the stable homogeneous (north pole)
solution may loose stability not only in a subcritical
{\sc Hopf} bifurcation (left to D) but also in a supercritical
one (between D and C).

In a neighbourhood of the codimension-2 point C there are
two completely different limiting cases concerning the
behaviour of the system.
The first one has to be
described (cf. region~IV) by the real equation
for the amplitude of the spatially periodic state,
the second one (cf. region~II)
by the complex equation --with all its implications--
governing the amplitude of the time-periodic mode.
Considering $\mu_a$ and $\mu_b$ as functions of the angle $\Phi$
the coupled system of
equations~(\ref{eqn:agl:codim})
provides a smooth transition between these dynamically radically
different situations.
There are furthermore saddle-node bifurcations
of spatially homogeneous limit-cycles on a line
emanating from D.
We will tackle this problem among other topics
in a subsequent paper.

Note that in the case of easy-plane anisotropy the
codimension-2 bifurcation always involves a subcritical
soft-mode bifurcation whereas it is supercritical
for easy-axis anisotropy.

%
%
\section{Summary}

The first goal of the present paper has been to discuss
systematically the instabilities of the spatially
homogeneous states of a strongly driven ferromagnet against
inhomogeneous perturbations.
Based on the work by~\cite{TTJS1995},
in which only homogeneous magnetizations were treated,
we have determined analytically the stability boundaries
of all the fixed point solutions in the spatially extended system.
Apart from the (spatially homogeneous) saddle-node and pitchfork bifurcations
we have found hard-mode ($\omega_c\neq 0,q_c=0$) and
soft-mode ($\omega_c=0, q_c\neq 0$) instabilities as well
as codimension-2 situations where both instabilities coincide.
The corresponding bifurcation lines have been presented in a
\dg parameter plane at fixed $h_\perp$ for
easy-axis and easy-plane anisotropy.

We then went on to derive the amplitude equations near those
instabilities which ultimately lead to pattern formation and
have calculated all of their coefficients up to third order
explicitly.
Their functional dependence on the system parameters was shown along each
bifurcation line which possesses at least one supercritical segment.
{}From this variation the locations in parameter space
where the real part of the third-order coefficients vanishes
have been determined for both types of instability.
At these points the bifurcation behaviour changes from sub- to supercritical.
For a {\sc{Hopf}} bifurcation  such a
degenerated situation is connected with a
saddle-node bifurcation of limit-cycles. The
respective bifurcation line in parameter space has been calculated
already in~\cite{TTJS1995}.
As explained in section~5.1 we expect an analogous
behaviour involving saddle-node bifurcations
of spatially periodic states in the soft-mode case.

For the interesting and more involved case of
the codimension-2 bifurcation we have found
two coupled Ginzburg-Landau equations, one with real
coefficients for the soft-mode amplitude and another one
with complex coefficients for the hard-mode amplitude.
This set of equations describes a smooth transition between two
fundamentally different limiting cases.
The first one is given by the relaxational dynamics of the
real Ginzburg-Landau equation for the amplitude of the spatially
periodic state.
The second one consists of the spatio-temporal
dynamics described by the complex Ginzburg-Landau equation governing
the amplitude of the time-periodic mode.
In an
intermediate range (cf. region (III) of figure~9) a competition
between both behaviours will occur.
A pretty complex situation, namely a bifurcation of even higher
codimension, arises if the value of $h_\perp$ is chosen
in such a way that the degenerated {\sc Hopf} bifurcation
falls together with the codimension-2 point.

We emphasize that it was the explicit though laborious determination
of the coefficients as functions of the physical parameters
which has revealed the existence of these rather intriguing situations.
As stated in the text there are other ones deserving closer attention.
We just mention two of them. The first one is the saddle-node bifurcation
of spatially periodic solutions, which is to be expected in the
vicinity of the degenerated soft-mode bifurcation.
The second one arises at the end points of our bifurcation lines.
At these points, representing further examples of bifurcations
of higher codimension, the perturbation theoretical ansatz
including the scaling of space and time will have to be modified
substantially.
We are working on these problems, hoping that our results will be
interesting not only from a more principal
point of view, but --in revealing new aspects of pattern formation
in magnetically ordered substances-- also under realistic experimental
conditions.

\section*{Acknowledgement}

The authors would like to thank Dr.~W.~Just for several fruitful and
illuminating discussions.
This work was performed within
a program of the Sonderforschungsbereich 185 Darmstadt--Frankfurt, FRG.

%
%
\section*{Appendix A: Derivation of the envelope equations\label{app:agln}}

This appendix contains the calculations of the explicit
expressions for the coefficients of the amplitude equations
starting from the basic equation of motion~(\ref{eqn:LLstereo}).
Considering their extent they are confined to a reasonable measure.
As a general reference for the procedure applied see for example
\cite{CrHo1993,KuTs1975,Mann1990}.
For the purpose of this section it is convenient to
formally rewrite the equation of motion using the
substitution $\Phi = \Phi_0 + \delta\Phi$ and
separate the linear and nonlinear contributions as
\be\label{eqn:formal}
  \hat\Gamma\,\delta\Phi = \imG \ N[\delta\Phi].
\ee
$N[\delta\Phi]$ contains all the nonlinear terms in $\delta\Phi$
while $\hat\Gamma$ which is given by \eqn~(\ref{eqn:LinStab})
collects the linear ones. Both of them depend
on the fixed point $\Phi_0$.

As has been exposed in section~\ref{sec:agln}
we introduce a small parameter~$\epsilon$
in order to study the system in the neighborhood
of the marginal stability line by writing the
parameters near their critical values ($\delta_c, \gamma_c$) as
\bea
  \delta & = & \delta_c + \epsilon^2 \,\delta_2 \\
  \gamma & = & \gamma_c + \epsilon^2 \,\gamma_2.
\eea
The solution of \eqn~(\ref{eqn:formal})
may then be assumed to be of the form
\be\label{eqn:exp:amplitude}
 \delta\Phi = \epsilon \ \Phi_1 + \epsilon^2 \ \Phi_2 + \epsilon^3 \ \Phi_3.
\ee
The function $\Phi_1$ contains the slowly varying amplitude~$A(X,T)$
together with the basic mode becoming unstable.
In the case of a soft-mode instability, for example, it is given by
\be
  \Phi_1 = v_0\ (A(X, T)\,e^{i\,q_c\,x}+ c.c.),
\ee
where $v_0$ is a complex number of modulus 1.
Considering first the linear part of \eqn~(\ref{eqn:formal})
several expansions have to be performed:
Because the fixed point coordinates $\Phi_0$ depend on the parameters
$\delta$ and $\gamma$ via \eqn~(\ref{eqn:fpSz})
we have to expand them with respect to $\epsilon^2$
\be
 \Phi_0 = \Phi_{00} + \epsilon^2\,\Phi_{02}.
\ee
The second order contributions are determined by
\be
 \left.\frac{df}{d\epsilon^2}\right|_{\epsilon=0} =
 \frac{\partial f}{\partial \Phi_0} \,\Phi_{02} +
 \frac{\partial f}{\partial\bar\Phi_0} \,\bar\Phi_{02} +
 \frac{\partial f}{\partial\delta} \delta_2+
 \frac{\partial f}{\partial\gamma} \gamma_2 = 0,
\ee
where $f$ is an abbreviation for the bracket on the right hand side of
\eqn~(\ref{eqn:LLstereo}),
disregarding the spatial derivatives.
This leads to
\bea
  \Phi_{00} \,\left( \delta_2 + i \,\gamma_2  \right) +
  \Phi_{02} \,\left( \delta_c  + i \,\gamma_c +
  h_\perp \Phi_{00} \right) \ + & & \\ \nonumber
  a \ \frac{\Phi_{02} \left(1 - 2\, | \Phi_{00} |^2 -
      |\Phi_{00}|^4 \right) -
      2 \,\bar\Phi_{02} \Phi_{00}^2 }{(\eppoo)^2}
  & = & 0.
\eea
The coefficients $\{\gs_1, \ldots,\gs_4\}$
appearing in the operator $\hat\Gamma$
depend on the parameters as well as on the fixed point coordinates,
which themselves are functions of these parameters.
Thus the second order quantities in
\be
  \gs_\alpha = \gs_\alpha^{(0)} +
                  \epsilon^2 \gs_\alpha^{(2)}, \hspace{2cm}
  \alpha = 1, \ldots, 4,
\ee
are found to be:
\bea
  \gs_1^{(2)} & = & \delta_2 + h_\perp\,\Re(\Phi_{02}) -
        \frac{4\,a}{(\eppoo)^3}\ \left(
        \pooq \Phi_{02} + \poo \bar{\Phi}_{02} \right) \\
  \gs_2^{(2)} & = & \gamma_2 + h_\perp\,\Im(\Phi_{02}) \\
  \gs_3^{(2)} & = & \frac{4\,a}{(\eppoo)^3}\
        \Re\left( -\poo\,\Phi_{02} +
        \poo^3\,\bar{\Phi}_{02}\right)\\
  \gs_4^{(2)} & = & \frac{4\,a}{(\eppoo)^3}\
        \Im\left( -\poo\,\Phi_{02} +
        \poo^3\,\bar{\Phi}_{02}\right).
\eea
Finally the introduction of slow space $X=\epsilon\,x$ and
time $T=\epsilon^2\,t$ coordinates leads to the substitutions
\be
  \partial_x \rightarrow \partial_x + \epsilon\,\partial_X, \hspace{1.5cm}
  \partial_t \rightarrow \partial_t + \epsilon^2\,\partial_T.
\ee
Putting all this together we obtain
\be\label{eqn:exp:linop}
 \hat\Gamma = \hat\Gamma_0 + \epsilon\,\hat\Gamma_1 +
              \epsilon^2\,\hat\Gamma_2,
\ee
with
\bea
  \hat\Gamma_0 &=& \partial_t - \imG \ \left( \gs_1^{(0)} - \partial_x^2 +
             i \,\gs_2^{(0)} +
            (\gs_3^{(0)} + i \,\gs_4^{(0)}) \ \conjC \right) \\
  \hat\Gamma_1 &=& €2\,\imG\,\partial_x\,\partial_X \\
  \hat\Gamma_2 &=& \partial_T - \imG \ \left(
              \gs_1^{(2)} + i \,\gs_2^{(2)} +
             (\gs_3^{(2)} + i \,\gs_4^{(2)}) \  \conjC \right).
\eea
Similarly the nonlinear part of \eqn~(\ref{eqn:formal})
can be expanded as:
\be\label{eqn:exp:nl}
  N[\delta\Phi] = \epsilon^2 \ N_2 + \epsilon^3 \ N_3
\ee
with
\bea
  N_2 &=& - \frac{2\, a}{(1+| \Phi_{00} |^2)^3}
        \left[ \bar\Phi_{00} \Phi_1^2 + 2\,\Phi_{00} |\Phi_1|^2 -
            \Phi_{00}^3 \bar\Phi_1^2 \right] +
            \frac{h_\perp}{2}\, \Phi_1^2 +
             \frac{2 \,\bar\Phi_{00}}{1+| \Phi_{00} |^2}\
               \left(\frac{\partial\Phi_1}{\partial x}\right)^2 \\
  N_3 &=& -\,\frac{4\,a}{(\eppoo)^3}
       \left[ \bar\Phi_{00} \Phi_1 \Phi_2 +
          \Phi_{00} (\bar\Phi_1 \Phi_2 + \Phi_1 \bar\Phi_2) -
          \Phi_{00}^3 \bar\Phi_1 \bar\Phi_2 \right] +
           h_\perp \Phi_1 \Phi_2 + \\ \nonumber
 & & \frac{2\, a}{(\eppoo)^4}
       \left[ (2\, \poobq - 1) \bar\Phi_1 \Phi_1^2 + \bar\Phi_{00}^2 \Phi_1^3 +
              3\, \Phi_{00}^2 \Phi_1 \bar\Phi_1^2 -
          \Phi_{00}^4 \bar\Phi_1^3 \right] + \\ \nonumber
 & & \frac{4\, \pooq}{\eppoo}\,\frac{\partial \Phi_1}{\partial x}\,
      \left(\frac{\partial\Phi_1}{\partial{}X}+
       \frac{\partial \Phi_2}{\partial{}x}\right) +
      2\,\frac{\bar\Phi_1 - \bar\Phi_{00}^2 \Phi_1}{(\eppoo)^2}
         \left(\frac{\partial\Phi_1}{\partial x}\right)^2.
\eea
Inserting the
expansions~(\ref{eqn:exp:amplitude},\ref{eqn:exp:linop},\ref{eqn:exp:nl})
into \eqn~(\ref{eqn:formal}) yields in ascending orders of $\epsilon$:
\bea
  \hat\Gamma_0 \, \Phi_1 &=& 0 \label{eqn:ordng1}\\
  \hat\Gamma_0 \, \Phi_2 &=& - \hat\Gamma_1 \, \Phi_1 +
                            \imG \ N_2 \label{eqn:ordng2}\\
  \hat\Gamma_0 \, \Phi_3 &=& - \hat\Gamma_2 \, \Phi_1 -
                               \hat\Gamma_1 \, \Phi_2 +
                             \imG \ N_3. \label{eqn:ordng3}
\eea
Observe that $\hat\Gamma_0$, containing a linear and an antilinear
part, is a linear operator when acting on real-valued functions.
In order to solve this hierarchy of equations we
notice that the solution of the first equation is
given by the critical mode itself (neglecting transient contributions).
This implements a certain definite wavenumber and/or frequency
into the theory and as a consequence
the right hand side of these equations,
depending on powers of this fundamental solution,
can be expanded with respect to harmonics of this mode.
In this way the solution of the problem may be constructed
within a space
of strictly periodic functions in time and/or space.
It is established by applying conventional Fredholm theory
which necessitates the introduction of an appropriate
scalar product.
An apt choice proves to be
\be\label{eqn:skal}
 (f,g) := \Re\left(\frac{1}{\Lambda_c}
        \int_0^{\Lambda_c} dx\  \frac{1}{T_c}\, \int_0^{T_c} dt
        \,\bar{f}(x,t) g(x,t) \right)
\ee
with $T_c = \frac{2\, \pi}{w_c}$ and $\Lambda_c = \frac{2\, \pi}{q_c}$.
For a soft-mode instability the critical mode is time
independent and \eqn~(\ref{eqn:skal}) reduces to an integration
over $x$, whereas for
a hard-mode instability we have to deal only with
a time-integration because there is no (fast) spatial coordinate.

In all three cases to be discussed the solvability condition
explained below is satisfied automatically
for \eqn~(\ref{eqn:ordng2}).
The amplitude equations are obtained from
the third-order \eqn~(\ref{eqn:ordng3})
which is solvable only if
\be \label{eqn:SolvCond}
  0 = -(\Psi_0, \hat\Gamma_2\,\Phi_1\,) - (\Psi_0, \hat\Gamma_1\,\Phi_2) +
          (\Psi_0, \imG \ N_3\,).
\ee
$\Psi_0$ is the left-null-eigenvector of $\hat\Gamma_0$
(which is of course the same as the right-null-eigenvector of
$\hat\Gamma_0^+$).
This condition, also known as the Fredholm alternative,
assures that the right hand side of \eqn~(\ref{eqn:ordng3})
contains no contribution exciting an eigenvector
of $\hat\Gamma_0$ with zero eigenvalue.
The ensuing construction guarantees,
that no secular terms occur in the perturbation expansion.
For further details see \cite{CrHo1993} and appendix A.2
of~\cite{Mann1990}.

We now split up our discussion
which up to now has been quite general according to
the different instabilities.

%
%
\subsection*{A.1 Soft-mode instability}

{}From \eqn~(\ref{eqn:ordng1}) the critical mode
is given by
\bea \label{eqn:crit:softmode}
  \Phi_1 &=& v_0 \ (A\,e^{i\,q_c\,x} + c.c.), \\ \label{eqn:soft:v0}
  v_0 &=& \frac{\gs_3^{(0)}+ i \,(\gs_2^{(0)}+\gs_4^{(0)})}
               {\sqrt{(\gs_3^{(0)})^2 + (\gs_2^{(0)}+\gs_4^{(0)})^2}}.
\eea
The left-0-eigenvector of $\hat\Gamma_0$ reads
\bea
  \Psi_0 &=& w_0 \, (D e^{i\,q_c\,x} + c.c.) \\
  w_0 &=& \imG\,\frac{\gs_2^{(0)}+\gs_4^{(0)} - i \,\gs_3^{(0)}}
        {\sqrt{(\gs_3^{(0)})^2 +  (\gs_2^{(0)}+\gs_4^{(0)})^2}}
\eea
where the (arbitrarily chosen) normalization
\be
 (w_0, v_0) = 1
\ee
proves to be convenient.
The complex parameter $D$ represents an undetermined
phase factor and is used later on for the extraction of the
amplitude equation.

Inserting $\Phi_1$ into the second order equation,
the latter can be solved because there are no resonant terms.
We get
\be
  \Phi_2 = 2\,c_{20}\,|A|^2  + c_{22}\,(A^2 e^{2\,i\,q_c\,x} + c.c.) -
           v_1\,\frac{1}{\gs_2^{(0)}}\,\partial_x\,\partial_X\,
           (A\,e^{i\,q_c\,x} + c.c.)
\ee
with
\bea
  v_1 &=& \frac{1}{\sqrt{(\gs_3^{(0)})^2 +
       (\gs_2^{(0)}+\gs_4^{(0)})^2}}\,
       \left(\gs_2^{(0)}+\gs_4^{(0)} - i\,\gs_3^{(0)} \right) \\
  c_{20} &=& \frac{1}{q_c^4} \,
      \left[ (q_c^2 + i \,\gs_2^{(0)}) \, (n_{2a} + n_{2b}) +
             (\gs_3^{(0)} + i\,\gs_4^{(0)}) \,
                (\bar{n}_{2a} + \bar{n}_{2b}) \right] \\
  c_{22} &=& \frac{1}{9\,q_c^4} \,
      \left[ (-3\, q_c^2 + i\,\gs_2^{(0)}) \, (n_{2a} - n_{2b}) +
             (\gs_3^{(0)} + i\,\gs_4^{(0)}) \,
                (\bar{n}_{2a} - \bar{n}_{2b})  \right] \\
  n_{2a} &=& \frac{h_\perp}{2}\, v_0^2 -
       \frac{2\,a}{(\eppoo)^3}
       \left[ \pooq\,v_0^2 + 2\,\poo\,|v_0|^2 - \poo^3\,\bar{v}_0^2 \right] \\
  n_{2b} &=& \frac{2\,\pooq\,v_0^2\,q_c^2}{\eppoo}.
\eea
There is an additional, arbitrary contribution proportional to $\Phi_1$
which belongs to the nullspace of the linear operator $\hat\Gamma_0$.
It is not considered any further because it does not
influence the amplitude equation.

Putting $\Phi_1$ and $\Phi_2$ into the third order \eqn~(\ref{eqn:ordng3}),
we finally can exploit the solvability condition~(\ref{eqn:SolvCond}).
Gathering all terms proportional to $\bar{D}$
we end up with a Ginzburg-Landau equation
with real coefficients:
\be\label{agl:softmode}
  \partial_T\,A = \mu_a\,A + \alpha_a\,\partial_X^2\,A - r_a \ |A|^2\,A.
\ee
Analogously the terms multiplying $D$ yield
the complex conjugate equation.
The coefficients are given explicitly by
\bea
 \mu_a    &=& - \, \frac{\epGq}{\gs_2^{(0)}} \,
   \left[ \gs_2^{(0)} \gs_2^{(2)} - \gs_3^{(0)} \gs_3^{(2)}
        - \gs_4^{(0)} \gs_4^{(2)} \right] \\
 \alpha_a &=& 2 \, \frac{\epGq}{\gs_2^{(0)}} \, q_c^2 \\
 r_a &=& \frac{\epGq}{\sqrt{(\gs_3^{(0)})^2 +
        (\gs_2^{(0)}+\gs_4^{(0)})^2}} \
        \Re\left( (\gs_2^{(0)}+ \gs_4^{(0)} + i \, \gs_3^{(0)}) \,
          (n_{3a}+n_{3b}) \right).
\eea
with
\bea
  n_{3a} &=& h_\perp\,v_0\,\left( 2\,c_{20} + c_{22} \right) - \\ \nonumber
 &&  \frac{4\,a}{(\eppoo)^3} \left[ (\pooq\,v_0+\poo\,\bar{v}_0) \,
     (2\,c_{20}+c_{22}) + (\poo\,v_0-\poo^3\,\bar{v}_0) \,
     (2\,\bar{c}_{20}+\bar{c}_{22}) \right] + \\ \nonumber
 && \frac{6\,a}{(\eppoo)^4}\, \left[ (2\,|\poo|^2 - 1)\,\bar{v}_0 v_0^2 +
     \pooq^2\,v_0^3+3\,\poo^2\,v_0\,\bar{v}_0^2-\poo^4\,\bar{v}_0^3 \right] \\
  n_{3b} &=& \frac{8\,\pooq\,v_0\,q_c^2}{\eppoo}\,c_{22} +
     \frac{2\,q_c^2\,v_0^2}{(\eppoo)^2} \,(\bar{v}_0 - \pooq^2 v_0).
\eea

%
%
\subsection*{A.2 Hard-mode instability}

The derivation of the amplitude equations for the
hard-mode instability proceeds along the same lines
as that for the soft-mode case:
the space-periodic critical mode has to be replaced merely
by a time-periodic one.
The main difference between the two instabilities is
caused by the symmetries of the basic equation of motion.
For a soft-mode instability the spatial inversion symmetry
leads to a Ginzburg-Landau equation with
real coefficients.
In the hard-mode case complex coefficients are admissible
because \eqn~(\ref{eqn:LLstereo}) is not
invariant under time reversal.

The linear mode at marginal stability is now given by:
\be
   \Phi_1 = \voa \    (B e^{i\,\omc\,t} + c.c.) +
            \vob \ i\,(B e^{i\,\omc\,t} - c.c.)
\ee
with
\bea\label{eqn:hard:v0a}
  \voa &=&\frac{1}{N_v}\,
       \left( \gs_3^{(0)} - \Gamma\,\gs_4^{(0)} - (\epGq)\,\gs_1^{(0)} +
       i\,(\gs_4^{(0)}+\Gamma\,\gs_3^{(0)}) \right) \\
  \label{eqn:hard:v0b}
  \vob &=&\frac{1}{N_v}\,i\,\omc\\
  N_v  &=&\sqrt{2\,(\epGq)\,\gs_1^{(0)}
      \,\left((\epGq)\,\gs_1^{(0)} +
        \Gamma\,\gs_4^{(0)}-\gs_3^{(0)}\right)}.
\eea
The left-0-eigenvector of $\hat\Gamma_0$ reads
\be
  \Psi_0 = \woa \    (E\,e^{i\,\omc\,t} + c.c.) -
           \wob \ i\,(E\,e^{i\,\omc\,t} - c.c.),
\ee
where
\bea
  \woa &=& \frac{1}{N_w}\
           \left( \gs_4^{(0)} + \Gamma\,\gs_3^{(0)} +
           i\,(-\gs_3^{(0)} + \Gamma\,\gs_4^{(0)} +
              (\epGq)\,\gs_1^{(0)})\right) \\
  \wob &=& - \frac{\omc}{N_w}\\
  N_w  &=& \frac{\omc}{2\,(\epGq)\,\gs_1^{(0)}} \ N_v,
\eea
and the following normalization has been chosen:
\bea
  (\woa, \voa) = (\wob, \vob) &=& 0 \\
  (\woa, \vob) = (\wob, \voa) &=& 1.
\eea
As before the second order \eqn~(\ref{eqn:ordng2})
has no resonant contributions and can be solved:
\be
  \Phi_2 = c_{2n} \,|B|^2 + c_{2p2} \ (B\,e^{i\,\omc\,t} + c.c.) +
                             c_{2m2} \ i\,(B\,e^{i\,\omc\,t} - c.c.)
\ee
with
\bea
  c_{2n} &=& \frac{\epGq}{\omc^2}\ \left[  -\gs_1^{(0)} (\epiG) \ n_{2n} +
     (\gs_3^{(0)}+i\,\gs_4^{(0)}) \ \bar{n}_{2n} \right] \\
  c_{2m2} &=& \frac{-1}{3\,\omc^2} \
       \left[2\,\omc\,\imG\,n_{2p2} +
        (\epGq) \left(\,i\,\imG\,\gs_1^{(0)}\ n_{2m2}
                  \right.\right. + \\ \nonumber
  & &   \left.\left. (\gs_3^{(0)}+i\,\gs_4^{(0)})\,\bar{n}_{2m2}
                  \right)\right] \\
  c_{2p2} &=& \frac{-1}{3\,\omc^2} \
      \left[\,(\epGq)\ \left( \,i\,\imG\,\gs_1^{(0)}\,n_{2p2} +
       (\gs_3^{(0)}+i\,\gs_4^{(0)})\,\bar{n}_{2p2}
                  \right) \right.- \\ \nonumber
  & &  \left. 2\,\omc\,\imG\,n_{2m2} \right] \\
  n_{2n} &=&  h_\perp\,v_{0a}^2 + h_\perp\,v_{0b}^2 -
      \frac{4\,a}{(\eppoo)^3}\
      \left(\pooq\,v_{0a}^2 + 2\,\poo\,v_{0a}\,\bar{v}_{0a} -
      \right.\\ \nonumber
  & & \left.\poo^3\,\bar{v}_{0a}^2 + \pooq\,v_{0b}^2 +
      2\,\poo\,v_{0b}\,\bar{v}_{0b} - \poo^3\,\bar{v}_{0b}^2 \right)\\
  n_{2m2} &=&  h_\perp\,v_{0a}\,v_{0b} - \frac{4\,a}{(\eppoo)^3}\
      \left(\pooq\,v_{0a}\,v_{0b} + \poo\,\bar{v}_{0a}\,v_{0b} +
      \right.\\ \nonumber
  & & \left. \poo\,v_{0a}\,\bar{v}_{0b} -
    \poo^3\,\bar{v}_{0a}\,\bar{v}_{0b}\right)\\
  n_{2p2} &=& h_\perp\,\frac{v_{0a}^2-v_{0b}^2}{2}  -
      \frac{2\,a}{(\eppoo)^3}\ \left(\pooq\,v_{0a}^2 +
      2\,\poo\,v_{0a}\,\bar{v}_{0a} - \right.\\ \nonumber
  & & \left. \poo^3\,\bar{v}_{0a}^2 -  \pooq\,v_{0b}^2 -
      2\,\poo\,v_{0b}\,\bar{v}_{0b} + \poo^3\,\bar{v}_{0b}^2\right)
\eea
In agreement with the soft-mode case we have
dropped a term belonging to the
kernel of $\hat\Gamma_0$.
Using this solution the nonlinear contribution $N_3$
of \eqn~(\ref{eqn:ordng3}) can be expressed as
\be
 N_3 = |B|^2\ \left(n_{3p1}\,(B\,e^{i\,\omc\,t} + c.c.)+
                    n_{3m1}\,i\,(B e^{i\,\omc\,t} - c.c.)\right) +
  \sum_{\nu \neq \pm 1} n_{3\nu} \,e^{i\,\nu\,\omc\,t}.
\ee
We refrain from presenting the complete expressions for the
coefficients $n_{3p1}$ and $n_{3m1}$ because they are rather lengthy.
They are received by expanding the nonlinear part $N_3$
in terms of the real valued functions $(B\,e^{i\,\omc\,t} + c.c.)$
and $i\,(B\,e^{i\,\omc\,t} - c.c.)$.
Using them the solvability condition for the
third order \eqn~(\ref{eqn:SolvCond}) can be evaluated.
Again we obtain the amplitude equation by gathering all
terms in front of $\bar{E}$:
\be
  \partial_T\,B = (\mu_{bR} + i\,\mu_{bI})\,B +
                  \alpha_b\,(1 + i\,c_1)\,\partial_X^2\,B
                 - r_b\, (1 - i\,c_2) \ |B|^2\,B.
\ee
The coefficients of this complex Ginzburg-Landau equation
are:
\bea
  \mu_{bR} &=& -\gs_2^{(2)} - \Gamma\,\gs_1^{(2)} \\
  \mu_{bI} &=& \frac{\epGq}{\omc}
      \left( \gs_1^{(0)} (\gs_1^{(2)}-\Gamma\,\gs_2^{(2)})
             - \gs_3^{(0)}\,\gs_3^{(2)} -
             \gs_4^{(0)}\,\gs_4^{(2)} \right) \\
  \alpha_b &=& \Gamma \label{eqn:diffconst:hard}\\
  \alpha_b \cdot c_1 &=& - \frac{(\epGq)\,\gs_1^{(0)}}{\omc}\\
  r_b &=& -\frac{1}{2}\
        \left[(\woa, \imG\,n_{3m1}) + (\wob, \imG\,n_{3p1})\right] \\
  r_b \cdot c_2 &=& -\frac{1}{2}\
        \left[ (\woa, \imG\,n_{3p1}) - (\wob, \imG\,n_{3m1}) \right].
\eea

%

\subsection*{A.3 Codimension-2 point}

In this case the bifurcation
conditions~(\ref{eqn:BifBed:Soft}) and (\ref{eqn:BifBed:Hard}) for
a stationary~($\omega_c=0, q_c\neq 0$) and an
oscillatory~($\omega_c\neq 0, q_c = 0$) instability are satisfied
simultaneously.
This requirement distinguishes
a single point ($\delta_c, \gamma_c$)
in the parameter plane and implies the relations
\bea
   \omega_c^2 &=& (\epGq)\,q_c^4 \\
   \gs_1^{(0)} &=& - q_c^2 \\
   \gs_2^{(0)} &=&  \Gamma\,q_c^2.
\eea
The critical mode is now composed of two degenerated modes and consequently
one has to introduce two amplitudes $A(X,T)$ and $B(X,T)$ to describe
their slow modulations, respectively:
\be
 \Phi_1 = v_0 \ (A(X,T) e^{i\,q_c\,x} + c.c.) +
                v_{0a} \ (B(X,T) e^{i\,\omega_c\,t} + c.c.) +
                v_{0b} \ i (B(X,T) e^{i\,\omega_c\,t} - c.c.).
\ee
$v_0$, $v_{0a}$ and $v_{0b}$ are defined as previously, compare
\eqn{}s~(\ref{eqn:soft:v0},\ref{eqn:hard:v0a},\ref{eqn:hard:v0b}).
The fact that two independent
degenerated critical modes occur entails that two different
left-0-eigenvectors enter the formulation of the solvability condition
\bea
 \Psi_{0a} &=& w_0 \ (E e^{i\,q_c\,x} + c.c.)  \\
 \Psi_{0b} &=& w_{0a} \ (D e^{i\,\omega_c\,t} + c.c.) +
               w_{0b} \ i (D e^{i\,\omega_c\,t} - c.c.).
\eea
$E$ and $D$ are arbitrary complex parameters.
Because most of the following expressions are very lengthy,
we do not present them in detail but restrict ourselves
to just describing how to get them.
Using a software package for symbolic mathematics it is
straightforward to obtain them explicitly.

The notation being necessary to keep book
of all the different time and space periodic functions and their
coefficients is as follows:
Introducing a real-valued basic set of functions,
which consists of fourier modes and their harmonics
\bea
  A_{n+}(x) &=& A^n  e^{i\,n\,q_c\,x} + c.c. \\
  A_{n-}(x) &=& i\,(A^n  e^{i\,n\,q_c\,x} - c.c. ) \\
  B_{n+}(t) &=& B^n  e^{i\,n\,\omega_c\,t} + c.c.\\
  B_{n-}(t) &=& i\,(B^n  e^{i\,n\,\omega_c\,t} - c.c.),
     \hspace{2cm} n = \ 1, 2, 3, \ldots
\eea
all functions may be expanded with respect to them.
Of course $A=A(X,T)$ and $B=B(X,T)$ still depend
on the slow variables.
Therefore by inserting in a first step
the linear mode $\Phi_1$ into the second-order
\eqn~(\ref{eqn:ordng2})
its right hand side is built up by the host of functions
\be
 \{ |A|^2, \,|B|^2, \, A_{1-}, \, A_{2+}, \,
    B_{2+}, \,B_{2-}, \,A_{1+}\cdot B_{1+}, \,A_{1+}\cdot B_{1-} \}.
\ee
Up to an unimportant term belonging to the nullspace of $\hat\Gamma_0$
the solution of this equation proves to be
\bea
 \Phi_2 &=& c_{2A}(0,0)\ |A|^2 + c_{2B}(0,0)\ |B|^2 +
            c_{2A}(2,0) A_{2+} + c_{2B+}(02)\ B_{2+} + \\ \nonumber
  & & c_{2B-}(0,2)\ B_{2-} +
      c_{2AB+}(1,1)\ A_{1+}\,B_{1+} +
      c_{2AB-}(1,1)\ A_{1+}\,B_{1-} + \\ \nonumber
  & & c_{2}(1,0)\ \partial_x\,\partial_X\ A_{1+}.
\eea
Here the indices ($\mu,\nu$) of the coefficients $c_{2...}$
mark spatial and temporal fourier modes, respectively.
They are calculated easily by inverting the linear operator on the
sub-spaces spanned by our basic set of functions.
Most of these calculations have been performed already in
the sections dealing with the soft- and hard-mode instabilities.
Only terms made up of products of spatial and temporal fourier modes
require additional computations.

Putting $\Phi_1$ and $\Phi_2$ into the remaining third-order
\eqn~(\ref{eqn:ordng3}) $N_3$ becomes
\bea
 N_3 &=& n_{3A+}(1,0)\,|A|^2\,A_{1+} + n_{3B+}(1,0)\,|B|^2\,A_{1+} +
         n_{3B+}(0,1)\,|B|^2\,B_{1+} +\\
  & &    n_{3B-}(0,1)\,|B|^2\,B_{1-} +
         n_{3A+}(0,1)\,|A|^2\,B_{1+} + n_{3A-}(0,1)\,|A|^2\,B_{1-} + \ldots
\eea
where the dots stand for all those modes which are nonresonant because
they are either higher order harmonics or constants.
Evaluating the solvability condition pertaining to
\eqn~(\ref{eqn:ordng3}) for each of the
two left-0-eigenvectors separately we obtain by collecting
all terms belonging to $\bar{E}$ or $\bar{D}$, respectively,
the following equations governing the time-evolution
of the amplitudes:
\bea
  \partial_T\,A &=& \mu_a\,A + \alpha_a\,\partial_X^2\,A
                     - r_a \ |A|^2\,A - s_a \,|B|^2\, A  \\ \nonumber
  \partial_T\,B &=& (\mu_{bR} + i\,\mu_{bI})\,B +
                  \alpha_b\,(1+ i\,c_1)\,\partial_X^2\,B
                 - r_b\,(1 - i\,c_2) \ |B|^2\,B
                 - (s_{bR} + i\,s_{bI}) \ |A|^2\,B.
\eea
Hence we have found a Ginzburg-Landau equation with real coefficients
for the amplitude $A$
being coupled to a complex one for the amplitude $B$.
The coupling coefficients are
\bea
  s_a &=& -(w_0, \imG \,n_{3B+}(1,0)) \\
  s_{bR} &=& -\frac{1}{2}\ \left[
        (w_{0a}, \imG\,n_{3A-}(0,1)) +
        (w_{0b}, \imG\,n_{3A+}(0,1)) \right] \\
  s_{bI} &=& \frac{1}{2}\ \left[
       (w_{0a}, \imG \,n_{3A+}(0,1)) -
       (w_{0b}, \imG \,n_{3A-}(0,1)) \right].
\eea
All the other first and third-order coefficients are the same
as in the soft- and hard-mode cases as is clear from
continuation arguments.

%
%

%
%
%
\pagebreak
\noindent
\section*{Figure captions}
\begin{itemize}
\item[Fig.1]
Saddle-node bifurcation line of \eqn~(\ref{eqn:LLstereo})
in the \dg parameter plane at $\Gamma=0.1$ and $h_\perp=0.1$.
The number of spatially homogeneous fixed point solutions is indicated.
\item[Fig.2]
Bifurcation diagrams including the hard- and soft-mode instabilities
of \eqn~(\ref{eqn:LLstereo}) at $\Gamma=0.1$ and $h_\perp=0.1$,
(a) easy-axis, $a=+1$,
(b) easy-plane, $a=-1$.
The region near the homogeneous cusp point is magnified in the insert.
\item[Fig.3]
Spectra of the linearized operator at various points
of the bifurcation diagrams (cf.fig.~2).
(1) degenerated soft- and hard-mode point,
(2) homogeneous cusp point, (3) end point of the soft-mode line,
(4) homogeneous cusp point, (5) Arnold-Takens-Bogdanov point,
(6) homogeneous cusp point.
\item[Fig.4]
Coefficients of the soft-mode
amplitude equation for $a=+1$ on the
bifurcation line~$\overline{13}$ of fig.~2a.
The coefficient $r_a$ changes sign at
$\delta\simeq -0.80$, $\gamma\simeq 0.18$.
\item[Fig.5]
Schematic bifurcation diagram for
the degenerated soft-mode instability.
The labeling is explained in the text.
\item[Fig.6]
Coefficients of the hard-mode
amplitude equation for $a=+1$ on the lower bifurcation line
of fig.~2a.
The coefficients $c_1$ and $c_2$ are shown separately in the insert.
$r_b$ changes sign at $\delta\simeq -1.10$, $\gamma\simeq 0.036$.
\item[Fig.7]
Coefficients of the hard-mode
amplitude equation for $a=-1$,
(a) on the upper hard-mode line,
(b) on the lower hard-mode line
of fig.~2b.
The coefficients $c_1$ and $c_2$ are shown separately in the insert.
$r_b$ changes sign at $\delta\simeq -1.10$, $\gamma\simeq 0.036$.
The region near zero is displayed on a smaller scale
(of order $10^{-4}$).
\item[Fig.8]
Coefficients of the codimension-2 amplitude equations
(a) for $a=+1$,
(b) for $a=-1$.
The coefficients $c_1$, $c_2$ and $\alpha_a$
are shown separately in the insert.
$r_b$ changes sign at
(a) $h_\perp\simeq 0.90$, $\delta\simeq -2.05$, $\gamma\simeq 0.21$,
(b) $h_\perp\simeq 1.72$, $\delta\simeq -1.12$, $\gamma\simeq 0.31$.
\item[Fig.9]
Schematic bifurcation diagram of the codimension-2 bifurcation
and the degenerated hard-mode bifurcation at $a=+1$.
The labeling is explained in the text.
\end{itemize}
%
%
\end{document}